\documentclass[12pt]{article}

\usepackage{graphicx}
\usepackage{graphics}

\begin{document}

\newcommand{\comment}[1]{}
\begin{center}
{\large \bf  \boldmath
Polarized $Z$ cross sections in Higgsstrahlung for the determination of
anomalous $ZZH$ couplings
}

\bigskip\smallskip
Kumar Rao$^a$, Saurabh D. Rindani$^b$, Priyanka Sarmah$^a$, Balbeer
Singh$^c$

\bigskip\smallskip
{\it $^a$Department of Physics, Indian Institute of Technology Bombay,\\ Powai,
Mumbai 400076, India}

\smallskip

{\it $^b$Theoretical Physics Division, Physical Research Laboratory,\\
Navrangpura, Ahmedabad 380009, India
}

\smallskip

{\it $^c$Department of Theoretical Physics,\\ Tata Institute of
Fundamental Research,
Dr. Homi Bhabha Road, \\Colaba, Mumbai 400005, India}

\bigskip \smallskip

{\small \bf Abstract}

\end{center}
\smallskip

{\small
The production of a Higgs boson in association with a $Z$ at an
electron-positron collider is one of the cleanest methods for the
measurement of the couplings of the Higgs boson. In view of the large
production cross section at energies a little above the threshold, it
seems feasible to make a more detailed study of the process by measuring
the cross sections for polarized $Z$ in order to measure possible
anomalous $ZZH$ couplings. We show that certain combinations
of cross sections in $e^+e^- \to ZH$ with different $Z$ polarizations
help to enhance or isolate the effect of one of the two kinds of
anomalous $ZZH$ couplings possible on general grounds of CP and Lorentz 
invariance. 
These combinations can be useful to get information on the $ZZH$
coupling in the specific contexts of an effective field
theory, two-Higgs-doublet models, and composite Higgs models, in a
relatively model-independent fashion. We find in particular 
that the longitudinal helicity fraction of the $Z$ is expected to be
insensitive to anomalous couplings, and would be close to its value in
the standard model in the scenarios we consider.  
We also discuss the sensitivity of the proposed
measurements to the anomalous couplings, including longitudinal 
beam polarizations, which suppress backgrounds, and can improve the
sensitivity if appropriately chosen.
}

\bigskip\smallskip
%\section{Introduction}
Experiments at the Large Hadron Collider (LHC) have studied in great
detail the properties of the Higgs boson, especially its couplings to
fermions and gauge bosons, with increasing precision. The results seem to
be in agreement with the predictions of the standard model (SM) 
to a good degree of accuracy. However, there is still a possibility that
the Higgs couplings are not precisely those predicted by the SM, 
and more data
from future experiments at the LHC should be able to improve the
accuracy of the comparison. 

Another prospect for acquiring more information on the Higgs being
considered is the construction of an electron-positron collider. 
At such a collider, the process  of associated $Z$ and Higgs production 
would be an attractive way to study
Higgs properties because of a clean environment and a  reasonably large 
cross section at energies not far above threshold. In addition, 
the Higgs energy-momentum can be reconstructed from the $Z$ decay
products regardless of the Higgs decay final state.  

In the context of the process $e^+e^- \to ZH$ alluded to above, 
deviations from the SM predictions, if seen, would most likely indicate 
$ZZH$ couplings differing from those in the SM, and perhaps also $\gamma
ZH$ couplings. We concentrate here on so-called anomalous $ZZH$
couplings and their effect on the $ZH$ production process. Our aim is to
investigate how $Z$ polarization can be used to determine the 
coefficients of different Lorentz tensors in a general $ZZH$ vertex
written in a model-independent way. 

In \cite{Rao:2019hsp} we considered angular asymmetries of
charged leptons arising in $Z$ decay which characterize the $Z$ spin
density matrix and could be used to determine $ZZH$ couplings. These
included polar and azimuthal angle asymmetries of leptons, the azimuthal
angular dependence arising from the off-diagonal elements of the density
matrix, based on the formalism of 
\cite{Boudjema:2009fz, Rahaman:2017qql}.
Here we concentrate on the simpler diagonal elements of the
density matrix, which are the degrees of polarization of the $Z$, and 
which can be probed without a detailed study of the
azimuthal distribution of leptons. A significant result is that 
certain combinations of
the polarized $Z$ cross sections enable isolation of specific anomalous
couplings. 

Our study is in the context of a few specific scenarios for $ZZH$ 
couplings differing from those in the SM. Ref. \cite{Rao:2019hsp} 
dealt with a completely model
independent set of form factors, only constrained by Lorentz
invariance. While one could measure several asymmetries and use them
simultaneously to determine or limit the form factors, the process would
be fairly complicated, and would be lacking in accuracy because of the
many variables involved. 
However, in the special scenarios considered in this paper, using 
combinations of polarized cross sections, a certain conceptual and
practical simplification results, and the dependence of the observables
is only on one, or at most two parameters, leading to better accuracy.
Though these scenarios vary in the assumptions that are made in them, 
they are reasonably model-independent, within those set of assumptions.
The different combinations of polarized cross sections that we suggest 
have a varied advantage in each scenario.

In all cases, we assume that longitudinally polarized beams would be
available, whose signs can be chosen to suppress background and improve the
sensitivity of the observables.

Other studies involving $Z$ polarization are
\cite{Ots:2000pq},
in the context of the process $e^+e^- \to ZH$ and 
\cite{Rao:2020hel, Nakamura:2017ihk} in the context of 
determination of anomalous $ZZH$ couplings at the LHC.

There have been several studies in the past for the measurement of
anomalous $ZZH$ couplings using final state asymmetries in 
$e^+e^- \to ZH$ without
reference to $Z$ polarization, see \cite{ZZH, hagiwara} and references 
therein. 
These do not have the advantage of the
kind of observables we are suggesting here. 
%In the next section we describe how a general $ZZH$ coupling may be
%written and the scenarios we propose to discuss.

%\section{Possibilities for $ZZH$ couplings}
We consider the process $e^+e^-\rightarrow Z H$,
where  the vertex $Z^\ast_{\mu}(k_{1})\rightarrow Z_{\nu}(k_{2}) H$  
has the Lorentz structure
\begin{equation}\label{vertex}
 \Gamma^{V}_{\mu \nu} =\frac{g}{\cos\theta_{W}}m_{Z} \left[ a_{Z}g_{\mu 
\nu}+
\frac{ b_{Z}}{m_{Z}^{2}}\left( k_{1 \nu}k_{2 \mu}-g_{\mu \nu}  k_{1}. 
k_{2}\right) +\frac{\tilde b_{Z}}{m_{Z}^{2}}
\epsilon_{\mu \nu \alpha \beta} k_{1}^{\alpha} k_{2}^{\beta}\right]  ,
 \end{equation}
where $g$ is the $SU(2)_L$ coupling and $\theta_{W}$ is the weak mixing 
angle. The couplings $ a_{Z}$, $b_Z$ and $\tilde b_{Z}$  are 
Lorentz scalars, and
depending on the framework employed, are either real constants, or
complex, momentum-dependent, form factors. The $a_Z$ and $b_Z$ terms 
are invariant under  CP, while  the 
$\tilde b_{Z}$ term corresponds to CP violation. In the SM, at 
tree level, the coupling $ a_{Z}=1$, whereas the other two couplings 
$b_{Z}$ and $\tilde b_{Z}$ vanish.

We consider here some possibilities for the couplings $a_Z$, $b_Z$ and 
$\tilde b_Z$ in various scenarios. 

\noindent{\bf (a)} In an effective field theory
(EFT) description of new physics 
\cite{Buchmuller:1985jz, Grzadkowski:2010es} 
(for applications to $ZH$ production,
see the works of Hagiwara and Stong in \cite{hagiwara} and of Beneke et
al., in \cite{ZZH}), where the SM is the 
low-energy limit of an extended theory, 
$a_Z$ would be normalized in the SM to the value of 1 
and would get a contribution $\delta a_Z$ of order $1/\Lambda^2$ from  
dimension-six operators, so that $a_Z = 1 + \delta a_Z$.
$b_Z$ and $\tilde b_Z$ would get contributions of order $1/\Lambda^2$ 
from dimension-six operators, and would be suppressed. They would, 
however, be real, from Hermiticity.

The EFT Lagrangian, including terms up to dimension 6 takes the form
\begin{equation}
{\cal L}_{\rm eff} = {\cal L}_{\rm SM}^{(4)} + \frac{1}{\Lambda^2}
\sum^{59}_{k=1} \alpha_k {\cal O}_k,
\end{equation}
with a sum of 59 independent terms of dimension 6, of which 11 are
relevant for our process. Of these, we can identify 
$\delta a_Z = \hat \alpha^{(1)}_{ZZ}$ and 
$b_Z = \hat \alpha_{ZZ}$,
where $\hat \alpha^{(1)}_{ZZ}$ and $\hat \alpha_{ZZ}$ in the notation of
\cite{Beneke:2014sba} are combinations of coefficients of dimension-6 
operators with a 
weak coupling factor $m_Z^2(\sqrt{2}G_F)^{1/2}$ pulled out. 
%There would also be a contribution from $e^+e^-ZH$ contact interactions
%present in EFT. However, we do not consider it here.

\noindent{\bf (b)} In two-Higgs-doublet models (for a review, see \cite{branco}) 
at tree level, $a_Z= \sin(\alpha - \beta)$, 
where $\tan\beta = v_2/v_1$, $v_{1,2}$ being the vacuum expectation
values of the two neutral Higgs fields, and $\alpha$ is the mixing 
angle
between the physical scalar eigenstates and Lagrangian scalar fields.
$\alpha$
can have a real value different from 1, whereas $b_Z$ and
$\tilde b_Z$ are zero. 
In these models, we will consider the coupling of the lighter of
the two CP-even Higgs particles.
In the case of other extensions of the sector with more doublets, 
singlets or triplets, the situation would be
similar. 

\noindent{\bf (c)} In composite Higgs models \cite{composite1,composite2,composite3}
the coupling $a_Z$ is different from 
unity, modified by a model-dependent reduction factor.
This factor in the so-called minimal composite Higgs models  
\cite{ composite2, Georgi:1984af} is described by one parameter $\xi$, 
and is given by
$\sqrt{1-\xi}$, where $\xi=v^2/f^2$, $v$ being the scalar vacuum
expectation value characterizing the electroweak breaking scale, 
and $f$ being the scale of compositeness.
For $f$ of order TeV, $v/f \ll 1$, and 
$a_Z = \sqrt{1-\xi} \approx 1 -
\frac{1}{2} \xi$.
$b_Z$ in these models
is expected to be small, of order $m_Z^2/f^2$ \cite{composite3}. 

%\section{Combinations of polarized cross sections}

As mentioned earlier, we aim to constrain the above
anomalous couplings with specific combinations of polarized
cross sections.  A hermitian, traceless, $3\times  3$ density matrix 
gives a 
complete description of the polarization states of spin one, 
characterized by 8 polarization parameters \cite{Leader}.
The diagonal elements of the density matrix correspond to pure 
polarization
states. With the final-state phase space appropriately put in, 
these diagonal
elements would give us production cross sections with definite $Z$
polarization. We are interested in constructing combinations of 
these polarized $Z$ cross sections, such that each combination 
would be dominantly sensitive to one of the two couplings.

On the experimental side, polarization of weak gauge bosons has been 
measured at the LHC in  $W + \rm jet$ production
\cite{CMS:2011kaj,
ATLAS:2012au},
$Z + \rm jet$ production 
\cite{CMS:2015cyj, ATLAS:2016rnf},
$W$ produced in the decay of top quarks 
\cite{Aad:2020jvx}
and more recently in $WZ$ production
\cite{ATLAS:2019bsc}
and same-sign $WW$ production
\cite{CMS:2020etf}.
The gauge-boson polarizations and helicity fractions are 
inferred from the angular distributions of the fermions to which the
gauge bosons decay \cite{Stirling:2012zt}. It should be possible to
determine the $Z$ polarization in the $e^+e^- \to ZH$ in the 
same way. See also \cite{De:2020iwq} for a recent suggestion for measurement
of $W$ polarization.

For the following, we assume for simplicity of some of our expressions
that there is 
no CP violation, $\tilde b_Z = 0$. However, there is no loss of
generality because a nonzero $\tilde b_Z$ would
not contribute to the observables we consider here. The
implications on including loop-level contributions from, say,
triple-Higgs couplings can be different and important. 

The density matrix elements and the helicity amplitudes from which they
are constructed were obtained in \cite{Rao:2019hsp}. 
The relevant diagonal 
elements of the density matrix integrated over phase space 
are given by
%\begin{eqnarray}
\begin{equation}
 \sigma(\pm,\pm)
=\frac{F(P_L,\bar P_L)g^{4}m^{2}_{Z}\vert \vec k_Z \vert}{96\pi
\sqrt{s} \cos^{4}\theta_{W}(s-m_{Z}^{2})^2}
\left[\vert a_Z \vert^2
	- 2 {\rm Re}(a_Zb_Z^*) \frac{E_Z\sqrt{s}}{m_Z^2} 
	+ \vert b_Z \vert^2 \frac{E_Z^2 s}{m_Z^4} \right],
\end{equation}
\begin{equation}
\sigma(0,0)
=\frac{F(P_L,\bar P_L)g^{4}E^{2}_{Z}\vert\vec k_Z\vert}
{96\pi\sqrt{s}\cos^{4}\theta_{W}(s-m_{Z}^{2})^2}
\left[\vert a_Z \vert^2
	- 2 {\rm Re}(a_Zb_Z^*) \frac{\sqrt{s}}{E_Z} 
	+ \vert b_Z \vert^2 \frac{s}{E_Z^2} \right].
\end{equation}
Here,  
$E_Z = (s - m_H^2 + m_Z^2)/(2\sqrt{s})$
is the $Z$ energy in the c.m. frame, 
and $\vert \vec k_Z \vert = \sqrt{E_Z^2 - m_Z^2}$ is the  magnitude of
the $Z$ three-momentum.
$c_V$ and $c_A$ are respectively the vector and axial-vector couplings
of the $Z$ to the electron, given by
$c_V = \frac{1}{2}(-1+ 4 \sin^2\theta_W)$ and $ c_A = -\frac{1}{2}$.
The beam polarization dependence enters through the combination 
\begin{equation}
F(P_L,\bar P_L) = (1-P_L\bar P_L)(c_{V}^2+c_{A}^{2}-2P_L^{\rm
eff}c_Vc_A),
\end{equation}
where $P_L$ and $\bar P_L$ are respectively the degrees of electron and
positron beam longitudinal polarization, and 
$P_L^{\rm eff} = (P_L - \bar P_L)/(1 - P_L \bar P_L)$.
The total cross section is then $\sigma = \sigma(+,+)+\sigma(0,0) +
\sigma(-,-)$.
%\begin{eqnarray}
%\sigma 
%&=&\frac{F(P_L,\bar P_L)g^{4}\vert \vec k_Z \vert}{96\pi
%\sqrt{s} \cos^{4}\theta_{W}(s-m_{Z}^{2})^2}
%\nonumber\\
%&& \hskip -0.8cm 
%\times
%\left[\vert a_Z \vert^2(E_Z^2 + 2m_Z^2 )
%	- 6 {\rm Re}(a_Zb_Z^*) E_Z\sqrt{s} 
%%	+ \vert b_Z \vert^2 \frac{(2E_Z^2+m_Z^2) s}{m_Z^2} \right].
% 	+ \vert b_Z \vert^2 (2E_Z^2/m_Z^2+1) s \right].
%\end{eqnarray}
%\begin{equation}
%\sigma \!
%=\!\frac{F(P_L,\bar P_L)g^{4}\vert \vec k_Z \vert (E_Z^2+2m_Z^2)}{96\pi
%\sqrt{s} \cos^{4}\theta_{W}(s-m_{Z}^{2})^2}
%\!\left[\vert a_Z \vert^2
%\!	- \!\frac{6 {\rm Re}(a_Zb_Z^*) E_Zs^{\frac{1}{2}}}{E_Z^2+2m_Z^2} 
%%	+ \vert b_Z \vert^2 \frac{(2E_Z^2+m_Z^2) s}{m_Z^2} \right].
% 	\!+\! \frac{\vert b_Z \vert^2 (2E_Z^2+m_Z^2) s }{m_Z^2(E_Z^2 + 2m_Z^2)}\right].
%\end{equation}

We now take up three combinations constructed out of the 
polarized cross sections, and discuss their features and advantages.

\noindent {\bf 1.} The quantity $\sigma_T - 2 \sigma_L$, where $\sigma_T \equiv \sigma
(+,+) + \sigma (-,-)$ and $\sigma_L \equiv \sigma (0,0)$
are respectively the cross sections for the production of 
transverse and longitudinally
polarized $Z$, is independent of $b_Z$ to
first order, and depends only quadratically on $b_Z$.

In \cite{Rao:2019hsp} we showed that a decay-lepton angular 
asymmetry 
$A_{zz} = \frac{3}{16}\;(\sigma_T-2\sigma_L )/\sigma$
can be used to put a limit on Re~$b_Z$. 
However, the numerator of this asymmetry, which is proportional to 
\begin{equation}
\Delta \sigma_1 \equiv
\sigma_T  - 2 \sigma_L =
\frac{F(P_L,\bar P_L)g^{4}\vert \vec k_Z \vert^3}{48\pi
\sqrt{s} \cos^{4}\theta_{W}(s-m_{Z}^{2})^2}
\left[ -\vert a_Z \vert^2 + \frac{s}{m_Z^2} \vert b_Z \vert^2
\right],
\end{equation}
is actually independent of  $b_Z$, to first order in $b_Z$. 
The limit on Re~$b_Z$ from the asymmetry used in \cite{Rao:2019hsp} actually comes 
from the $b_Z$
dependence of the denominator, which is the cross section.
Thus, in models where $b_Z$ arises only as a small effect, possibly from
loops, the quantity $\sigma_T - 2\sigma_L$ 
can be used to determine $a_Z$ (assumed non-SM) independently of 
$b_Z$.

\noindent{\bf (a)} In EFT, with $a_Z = 1 + \delta a_Z$, and keeping only 
first order in higher-dimensional  couplings,
\begin{equation}
\Delta \sigma_1 =
\frac{F(P_L,\bar P_L)g^{4}\vert \vec k_Z \vert^3}{48\pi
\sqrt{s} \cos^{4}\theta_{W}(s-m_{Z}^{2})^2}
\left[-1 -2{\rm }\delta a_Z  
\right],
\end{equation}
and can be used to determine $\delta a_Z$.

\noindent{\bf (b)} In 2HDM like models, there is no tree-level 
higher derivative $b_Z$-type
coupling (which is non-renormalizable). 
$b_Z$ may arise only at loop level and is therefore small.
Hence the quadratic term in $b_Z$ can be neglected.
So $\Delta \sigma_1$ can be used
to determine $a_Z$, which is $\sin(\alpha-\beta)$, as mentioned earlier.

\noindent{\bf (c)}
Again, in the case of composite models, $b_Z$ is expected to be small,
and hence the quadratic term can be neglected.
Hence $\Delta \sigma_1$ can again be used to determine $a_Z$, and
hence the parameter $\xi$.

\noindent {\bf 2.} Another useful combination of polarized cross 
sections is
$\Delta \sigma_2 \equiv
\sigma_T - (2 m_Z^2/E_Z^2)  
\sigma_L$, which evaluates to
\begin{equation}
\Delta \sigma_2 
=\frac{F(P_L,\bar P_L)g^{4}\vert \vec k_Z \vert^3}{48\pi
\sqrt{s} \cos^{4}\theta_{W}(s-m_{Z}^{2})^2}
\left[- 2 {\rm Re} (a_Z b_Z^* )\frac{ \sqrt{s}}{ E_Z}
+   \displaystyle \vert b_Z \vert^2 \frac{ (E_Z^2 + m_Z^2)s}{E_Z^2m_Z^2} 
\right].
\end{equation}
This is independent of $|a_Z|^2$, and hence proportional to $b_Z$.

\noindent{\bf (a)}
To linear order in the EFT couplings,
\begin{equation}
\Delta \sigma_2 
=\frac{F(P_L,\bar P_L)g^{4}\vert \vec k_Z \vert^3}{48\pi
E_Z \cos^{4}\theta_{W}(s-m_{Z}^{2})^2}
\left(- 2 {\rm Re}~b_Z 
\right).
\end{equation}
$\Delta \sigma_2$ can thus be used to determine $b_Z$, thus giving
information complementary to that obtained from $\Delta \sigma_1$.

\noindent{\bf (b)}
In SM and models like 2HDM, $\Delta \sigma_2$ will be zero. Of
course, at the loop level, the answer will be non-zero, and this would
get dominant contribution from say triple-Higgs couplings, since
tree-level contributions are eliminated.

\noindent{\bf (c)} In composite models $\Delta \sigma_2$  will also be zero at tree
level, or in some models, suppressed by $m_Z^2/f^2$ \cite{composite3}.

\noindent{\bf 3.} The longitudinal helicity fraction  $F_0 \equiv \sigma_L/ \sigma$ 
is given by 
\begin{equation}
\frac{\sigma_L}{\sigma} = 
 \frac{E_Z^2 \vert a_Z \vert^2 - 2 {\rm Re}
(a_Zb_Z^*)E_Z\sqrt{s}+ s \vert b_Z \vert^2}{(2m_Z^2+E_Z^2) 
\vert a_Z \vert^2 - 6 {\rm Re} (a_Zb_Z^*) E_Z \sqrt{s} + s \vert b_Z 
\vert^2 (1+2 E_Z^2/m_Z^2)}.
\end{equation}
The value of the helicity fraction is independent of beam polarization.
To first order in $b_Z/a_Z$, 
\begin{equation}
F_0 = \frac{E_Z^2}{2m_Z^2 + E_Z^2}
 \left( 1  + \frac{{\rm Re}
(a_Zb_Z^*)}{\vert a_Z \vert^2} \frac{4\sqrt{s}
 \vert \vec k_Z \vert^2 
}{E_Z (2m_Z^2+E_Z^2) }
\right).
\end{equation}
\noindent{\bf (a)}
To first order in dimension-six EFT couplings,
\begin{equation}
F_0
\approx
\frac{E_Z^2}{2m_Z^2 + E_Z^2}
 \left( 1  + {\rm Re}
(b_Z)\frac{4\sqrt{s}
 \vert \vec k_Z \vert^2 
}{E_Z (2m_Z^2+E_Z^2) }
\right).
\end{equation}
We see that the SM contribution dominates. There is a correction 
from the $b_Z$ coupling, but no dependence on the $\delta a_Z$ type of
couplings. A measurement of the helicity fraction would thus be mainly
model independent, with only a mild dependence on the dimension-six
couplings entering $b_Z$.

\noindent{\bf (b)}
In models
like SM and 2HDM, where $b_Z = 0$, the helicity fraction 
is independent of $a_Z$  and therefore 
a model-independent quantity, $E_Z^2/(2 m_Z^2 + E_Z^2)$, which
approaches 1 at high energies \cite{Barger:1993wt}.
This high-energy behaviour is as expected because the longitudinally
polarized $Z$ bosons give a dominant contribution, which can be seen in eq.
(4).
The model-independence of $F_0$ seems interesting to verify 
experimentally. 
Corrections from loops would be interesting to check. 

\noindent{\bf (c)} In composite Higgs models, since $b_Z$ is expected to be small, the
value of $F_0$ is approximately the same as in the SM, or as
that 
in the 2HDM.

We now discuss how sensitive the measurements of
these three quantities would be in the determination of the relevant
couplings.

The statistical sensitivity is determined by comparing  the expected 
number of 
new-physics events $\Delta N$ with the statistical fluctuation 
$\sqrt{N_{\rm SM}}$ in the number of  events in the SM.
Thus, the 1-$\sigma$ limit on a coupling can be obtained by equating
these two quantities.
To linear order,
\begin{equation}
\Delta N_i = {\cal L}( \Delta \sigma_i- \Delta \sigma_i^{\rm SM}) 
= {\cal L}
\frac{\partial \Delta \sigma_i}{\partial c_j
} c_j = \sqrt{{\cal L} \sigma_i^{\rm SM}},
\end{equation}
where  $c_j$ ($j=1,2$) is the anomalous coupling ($\delta a_Z$ or $b_Z$)
and ${\cal L}$ is the integrated luminosity.
Then, the 1-$\sigma$ limit on $c_j$ from a measurement of $\Delta_i$ 
is given by 
\begin{equation}
c_j^{\rm lim} = \frac{\sqrt{\sigma_i^{\rm SM}}}
		{\sqrt{\cal L}\vert \partial \Delta \sigma_i/\partial c_j \vert}.
\end{equation}
  
At the LHC, a measurement of $W$ polarization in top decay with c.m.
energy of 8 TeV and an integrated luminosity of 20 fb$^{-1}$, an
accuracy of about 2\% for $F_0$ and about 3\% for  the left-handed
helicity fraction $F_L$ \cite{Aad:2020jvx}. This implies an overall
accuracy of 2-3\% for our observable $F_0$, though it is likely to 
be better for $Z$
polarization at an $e^+e^-$ collider. However, for the measurement of
our first two observables, which are combinations of polarized cross
sections and not ratios, the efficiency would be expected to be lower.
The efficiency would also depend on the actual final states into
which $Z$ and $H$ decay. The corresponding backgrounds and the
kinematic cuts needed to suppress the backgrounds would determine the
efficiency. A discussion of backgrounds for ealier proposals for linear
colliders may be found in \cite{djouadi} and references therein, 
and for more recently planned colliders can be
found in \cite{An:2018dwb} for the CEPC and \cite{ruan} for the 
FCC-ee and the CEPC. Here, we have not incorporated effects of cuts and
assumed an ideal efficiency.
We reiterate that our main motive is to show the usefulness of certain 
polarization combinations which have not been considered before.

\begin{figure}[b!]
\centering
\includegraphics[height=5.5cm]{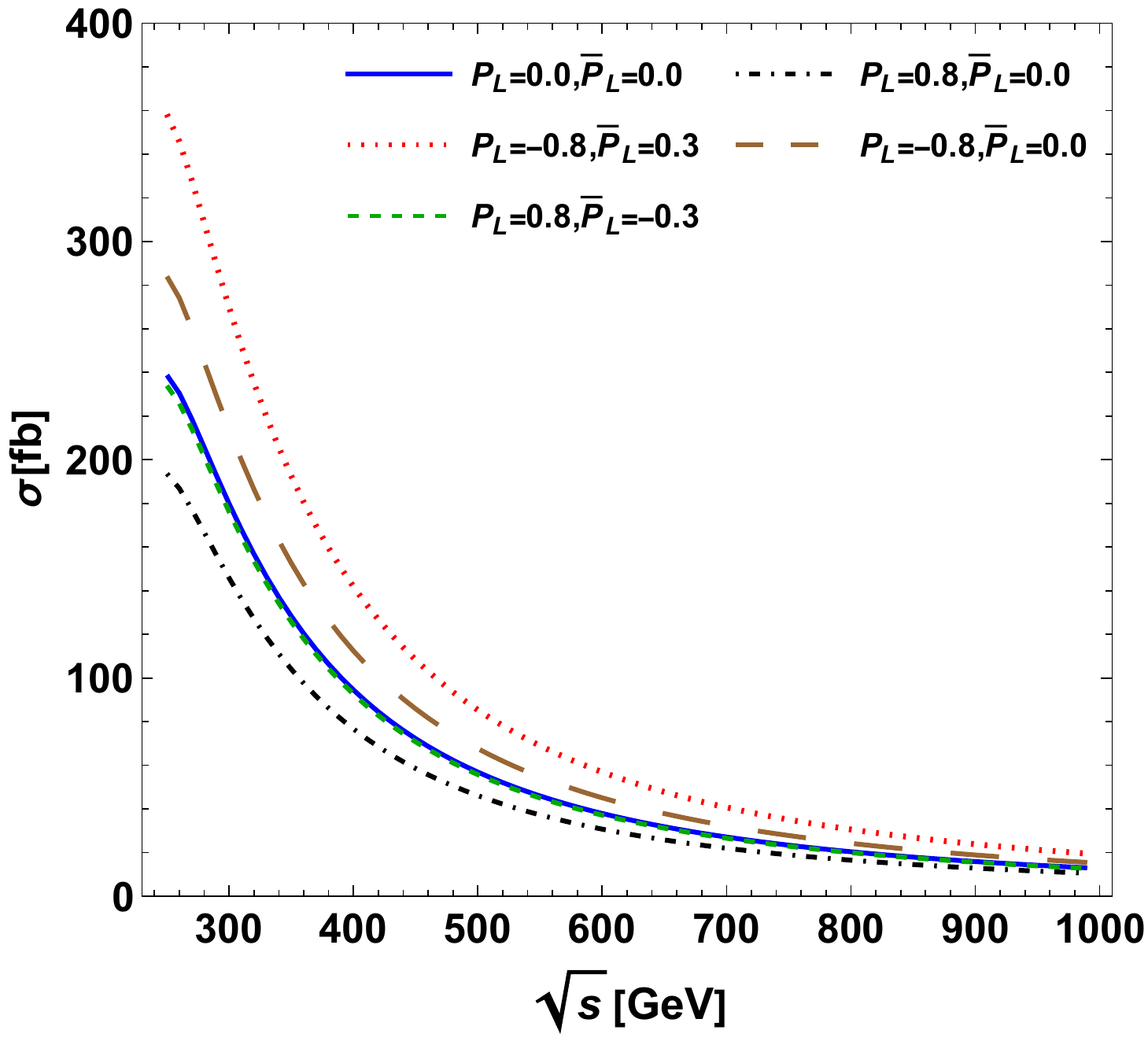}
\caption{The SM cross section $\sigma_{\rm SM}$ as a function
of $\sqrt{s}$ for different electron and positron polarization
combinations.
}
\label{sigma}\end{figure}
To get an idea of the event rates, we have first plotted in
Fig. \ref{sigma} the SM cross section as a function of the c.m. energy
for various polarization combinations.

We estimate the 1-$\sigma$ limits possible on $\delta a_Z$ using
$\Delta\sigma_1$ and on $b_Z$ using $\Delta \sigma_2$ 
for various configurations of $e^+e^-$ colliders, for both polarized and
unpolarized beams. We observe that there is improvement in the limits in the
presence of negative electron polarization and/or positive positron
polarization as compared to the unpolarized case. Furthermore, inclusion of
(negative) positron polarization results in an improvement as compared to
the case when only the electron beam is polarized. 
\begin{figure}[t!]
\centering
\includegraphics[height=5.5cm]{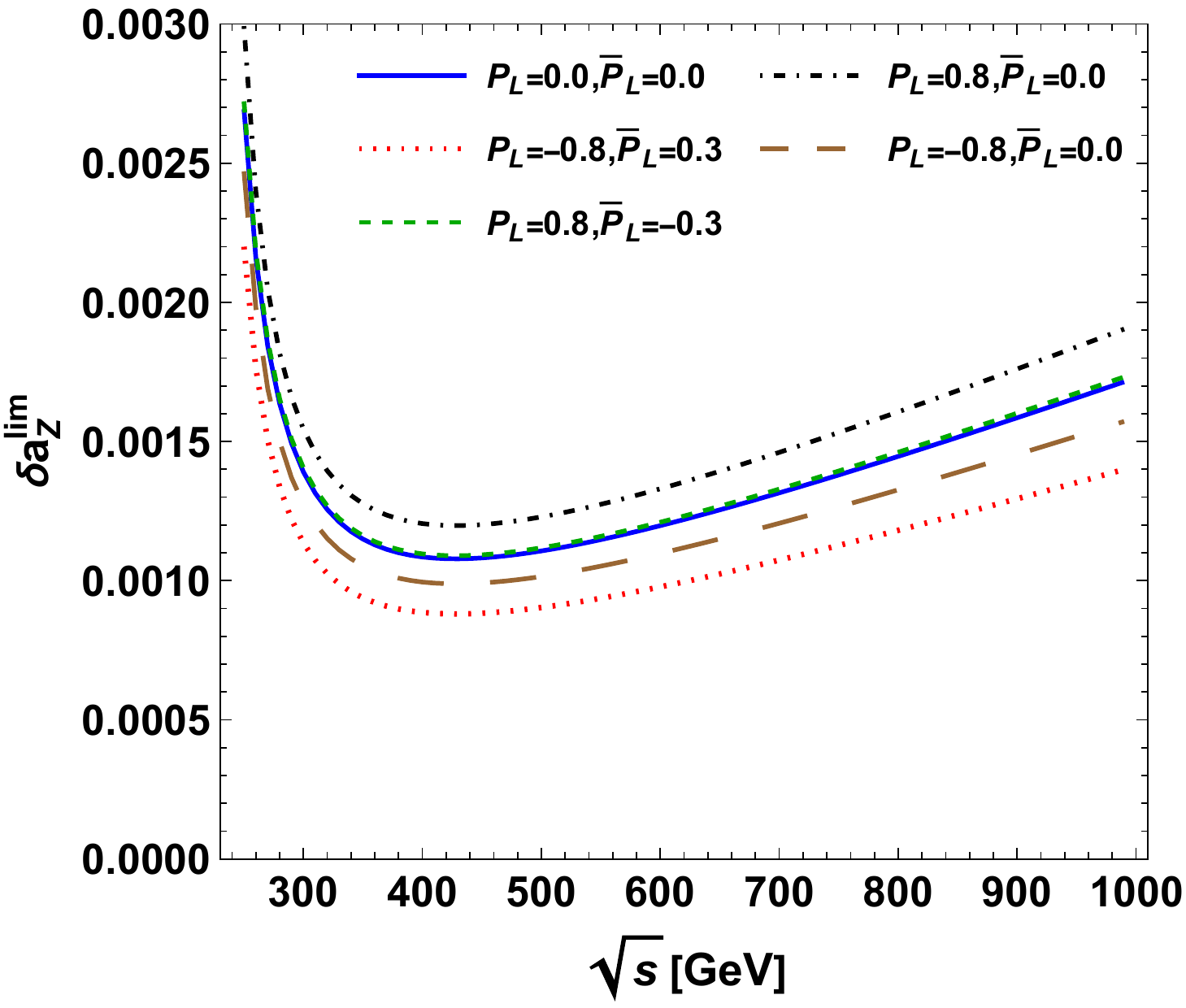}
\includegraphics[height=5.5cm]{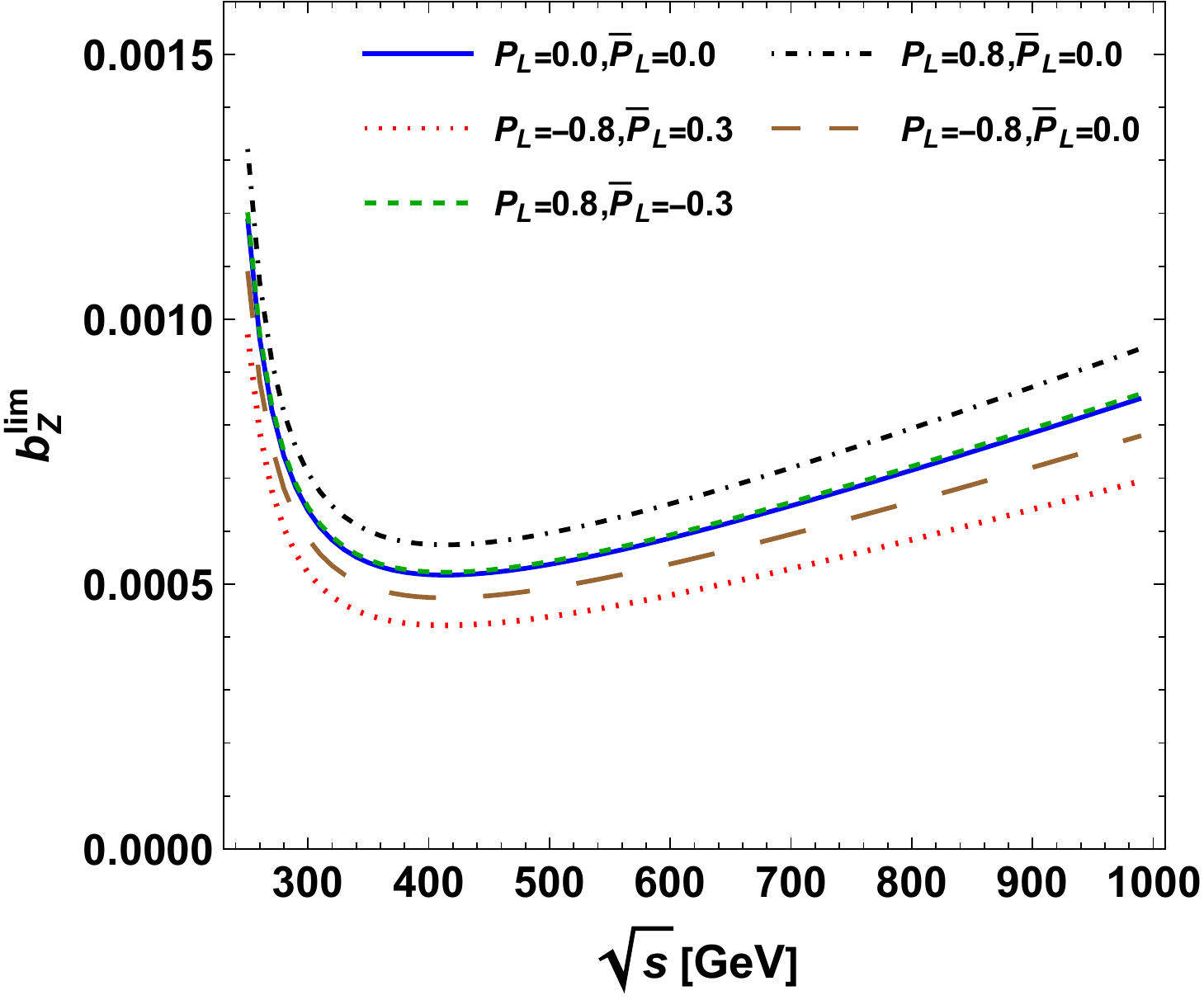}
\caption{
 1-$\sigma$ limits on $\delta a_Z$ (left panel) and on $b_Z$ (right
panel) which may be
obtained respectively from $\Delta\sigma_1$ and $\Delta\sigma_2$ 
for ${\cal L}= 2\, {\rm ab}^{-1}$
plotted 
as functions of $\sqrt{s}$ for different electron and positron polarization
combinations.
}
\label{azbzlimit}
\end{figure}

We have plotted in Fig. 2, 
for the case of ${\cal L}= 2\, {\rm ab}^{-1}$,
 the limits on $\delta a_Z$ and $b_Z$ as functions of the c.m. energy
for various combinations of beam polarization.
It is clear from these plots
that the optimal limits in either case are obtained for a
c.m. energy in the region of 350 GeV.
We also see that longitudinal beam polarization we have chosen helps to
improve the sensitivity to a great extent.
In addition to the results shown in the plots we find that further 
increase in positron polarization, if feasible, would improve the
sensitivity further.
As for example, for $\sqrt{s}=500$ GeV, in going from $P_{e^-} = -0.8$ 
and $P_{e^+} = +0.3$  to 
$P_{e^-} = -0.8$ and $P_{e^+} = +0.6$, the 
limit on $\delta a_Z$ improves from $9.1 \times 10^{-4}$ to $8.2 \times
10^{-4}$ and on $b_Z$ from  $4.4 \times 10^{-4}$ to $4.0 \times
10^{-4}$.

We may compare the limits we consider possible here using polarized
cross section combinations with the limits
estimated in \cite{Rao:2019hsp} with the use of asymmetries. The limits
on $b_Z$ we find here are better by an order of magnitude, 
even though in \cite{Rao:2019hsp} $a_Z$ was assumed to be exactly equal 
to 1, whereas here we only make a linear approximation.
It should of course be borne in mind that in either case, in practice, 
the limits would
be somewhat lower and dependent on a more detailed analysis including
the effects of cuts and detector efficiency, which we have not attempted
here.

%\section{Conclusions}
To conclude, we have studied how $Z$ polarization in $e^+e^- \to HZ$ 
can be used to measure
 general Lorentz-invariant $HZZ$ interaction described by three
couplings, $a_Z$, $b_Z$ and $\tilde b_Z$, of which the first two are CP
conserving. 
Instead of the full $Z$ spin density matrix 
which would require complicated distributions or asymmetries
of the $Z$ decay products, we suggest looking at only the diagonal 
elements of this matrix. These correspond to
cross sections for the production of $Z$ with  definite polarizations,
which are more easily accessible. We find that certain combinations of
$Z$ production cross sections with definite $Z$ polarization can 
help to enhance or isolate the effect of one of the two 
CP-even couplings. 

The specific combinations of longitudinally polarized ($\sigma_L$) and
transversely polarized ($\sigma_T$) cross sections we consider are
$\Delta \sigma_1 \equiv \sigma_T - 2\sigma_L$, 
$\Delta \sigma_2 \equiv
\sigma_T - (2 m_Z^2/ E_Z^2) 
\sigma_L$, and the longitudinal helicity fraction $F_0 \equiv
\sigma_L/(\sigma_L + \sigma_T)$.
% where $\sigma_L$ and $\sigma_T$ are respectively the
%cross sections with longitudinal and transverse polarizations of the
%$Z$, and $\sigma \equiv \sigma_L + \sigma_T$ is the total cross
%section. 
We show that $\Delta \sigma_1$ is independent of
$b_Z$ to first order, and can therefore be used to determine $a_Z$.
On the other hand,  $\Delta \sigma_2$ is proportional to $b_Z$ and
has no leading $|a_Z|^2$ dependence, so that it can be used to
determine $b_Z$.  To leading order in $b_Z/a_Z$, 
$F_0$ takes its SM value,
viz., $E_Z^2/(2 m_Z^2 + E_Z^2)$, regardless of the value of 
$a_Z$.

We consider in a relatively model-independent fashion three scenarios,
viz., EFT, 2HDM and composite Higgs, and analyze what role the
measurement of the three
cross section combinations $\Delta \sigma_1$, $\Delta \sigma_2$ and
$F_0$ can play in these scenarios. We also estimate the 1-$\sigma$ limits
on the couplings that could be placed using $\Delta \sigma_1$ and
$\Delta \sigma_2$.

We have also considered the effect of longitudinal beam polarization. 
While the limits on the couplings from $\Delta \sigma_1$ and 
$\Delta \sigma_2$ are improved by a choice of electron and positron beam
polarizations $-0.8$ and $+0.3$, $F_0$ is independent of beam
polarization.  

We find that limits of the order of a few
times $10^{-4}$ to $10^{-3}$ can be obtained for ranges of $\sqrt{s}$
from 250 to 500 GeV and 
${\cal L}$ values from 2 to 30 ab$^{-1}$ envisaged at various
electron-positron colliders.

The cross section combinations we have described here would be
interesting to  measure at future colliders.  A
more detailed analysis using event generators and detector simulation
would be necessary to check how well the estimates made here survive
after realistic kinematic cuts and detection efficiencies. 
Further, it would be interesting to investigate if such cross section
combinations yield useful results in a $pp$ environment, like at the
LHC.

\noindent {\bf Acknowledgement} The work of SDR was supported by the
Senior Scientist programme of the Indian National Science Academy, New
Delhi.

\end{document}